\PassOptionsToPackage{sort&compress}{natbib}
\documentclass[%
 reprint,
 superscriptaddress
]{revtex4-2}
 \usepackage{amsmath}
\usepackage{graphicx}
\usepackage{xcolor}
\usepackage{hyperref}
\usepackage[T1]{fontenc}
\usepackage[utf8]{inputenc}
\usepackage[sort&compress]{natbib}
\usepackage{xcolor} 

\begin{document}
\title{Inertial effects on fluid flow through low-porosity media}
\author{Sahrish B. Naqvi}%
 \email{sahrishbatool.naqvi@uwr.edu.pl}
 \affiliation
{
 Institute of Theoretical Physics, Faculty of Physics and Astronomy, University of Wroc{\l}aw, pl. M. Borna 9, 50-204 Wroc{\l}aw, Poland
}%
\author{Damian Śnieżek}
\affiliation
{
 Institute of Theoretical Physics, Faculty of Physics and Astronomy, University of Wroc{\l}aw, pl. M. Borna 9, 50-204 Wroc{\l}aw, Poland
}%
\author{Dawid Strzelczyk}
\affiliation
{
 Institute Jo\v{z}ef Stefan, Jamova cesta 39, 1000, Ljubljana, Slovenia
}%
\author{Mariusz Mądrala}
\noaffiliation
\author{Maciej Matyka}%
\affiliation
{
 Institute of Theoretical Physics, Faculty of Physics and Astronomy, University of Wroc{\l}aw, pl. M. Borna 9, 50-204 Wroc{\l}aw, Poland
}%


\date{\today}

\begin{abstract}
We investigated the nonlinear effects of gravity-driven fluid flow through a two-dimensional, low-porosity, packed bed of stubby stone grains. We focused on preferential channel formation, tortuosity, spatial distribution of kinetic energy,  and vortex formation. We show that nonlinear effects dominate at relatively high Reynolds numbers, even though the deviation from Darcy’s law is not visible in friction factor measurements. We further notice an increased flow asymmetry of the flow field revealed by vorticity analysis and surprising correlation between tortuosity and apparent permeability in the inertial flow regime. 
\end{abstract}

\maketitle

\section{Introduction}

Fluids, when flowing through porous media, form self-organized, preferential flow channels, which under high driving forces may be affected by inertial effects emerging at high Reynolds numbers \cite{andrade1999inertial, nissan2018inertial,sniezek2024inertia, matyka2017granica}. The dynamic interaction of the viscous fluid with the microstructure of the porous medium shapes preferential paths  \cite{hyman2020flow}. In a low-porosity medium, the fluid navigates through narrow tortuous pathways that follow the geometry of the pore space. The channeling effects are enhanced at low Reynolds number conditions, which may hinder the visibility of inertial effects in experiments \cite{wang2019experimental}. Understanding the mechanisms governing flow channelization in low-porosity media is essential across various domains including subsurface energy storage, CO$_2$ sequestration, and low-permeability reservoir engineering \cite{hovorka2019dynamic}. Low-porosity porous media are fundamental to various engineering applications, such as microfluidic device design \cite{cao2019application}, advanced energy systems \cite{banerjee2021developments}, high-efficiency filtration technologies \cite{zhang2024optimal}, and thermally active catalytic reactors \cite{wang2014thermal}, where constrained pore networks influence transport and reaction processes.

Traditionally, two main empirical models have been used to describe the flow through porous materials. In the laminar low-velocity regime, the Darcy law establishes a linear relationship between the forcing term and flow velocity:
\begin{equation}\label{eq:darcy}
\mathbf{U} = \frac{\kappa}{\mu} \textbf{f},
\end{equation}
where $\mathbf{U}$ is the superficial velocity, $\mu$ is the dynamic viscosity of the fluid, $\kappa$ is the medium's permeability, $L$ is the domain length, and $\textbf{f}$ the body force \cite{neuman1977theoretical,whitaker1986flow}. However, deviations from linearity are observed as the fluid velocity increases, and the inertial effects become significant. Thus, second-order correction is often introduced in the form of the Forchheimer equation:
\begin{equation}\label{eq:Forch}
\mathbf{U} = \frac{\kappa}{\mu}  \textbf{f} - \beta \rho \kappa^{1/2} |\mathbf{U}| \mathbf{U},
\end{equation}
where $\beta$ is an empirical inertial coefficient \cite{basak1977non,whitaker1996forchheimer,zeng2006criterion,arbabi2024transition}. The onset and strength of inertial transition depend sensitively on both the magnitude and spatial distribution of medium properties (e.g., porosity), that is, on a particular form of heterogeneity \cite{fourar2005inertia}. 
An increase in the driving forces governs the transition from the linear Darcy regime to the inertial flow, which comprises many physical phenomena. In particular, as the Reynolds number ($Re$) increases, local flow separation is observed with vortex formation and enhanced drag, resulting in the quadratic correction term in Eq. ~\ref{eq:Forch}   \cite{whitaker1996forchheimer,zeng2006criterion}. However, experimental studies have demonstrated that at very low Reynolds numbers, the flow may deviate from the linear behavior predicted by the Darcy law, commonly referred to as the pre-Darcy regime, where the pressure–velocity relationship becomes nonlinear even in the absence of inertial effects \cite{wang2019experimental}. Numerical simulations further support that the porosity is a decisive factor in determining the critical transition point to nonlinearity \cite{arbabi2024transition}. For example, pore-scale simulations of bead-packs, Bentheimer sandstone, and Estaillades carbonate indicate that despite having similar porosities, heterogeneous rocks such as Estaillades exhibit the onset of inertial flow at Reynolds numbers two to three orders of magnitude lower than in homogeneous media \cite{muljadi2015impact,muljadi2016impact}. These differences are attributed to the influence of the pore geometry and the formation of steady vortices that appear along with changes in the velocity distribution in the pore space. 

In a highly porous medium, the fluid gains momentum easily, enabling the early onset of nonlinearity and pronounced inertial effects \cite{andrade1999inertial}. However, these observations primarily stem from idealized or high-porosity systems. Natural porous materials, particularly those with low porosity, exhibit more intricate geometries of the pore space, where the structural heterogeneity and geometric tortuosity of pores play a dominant role and lead to vortex formation even at low Reynolds numbers. Our understanding of the low-porous medium is that it exhibits complex flow paths driven mainly by geometry. They are highly resistant, which weakens the inertial effects in flow \cite{arthur2020piv}. Thus, the physical mechanism of inertia and transition to nonlinearity is much more complicated than in a highly porous medium reported in \cite{andrade1999inertial,sniezek2024inertia}.

We aim to investigate real, low-porosity geological rock sample to understand better the transition from a linear to a nonlinear regime with increasing Reynolds number. We explore how nonlinearity emerges in a complex, low-porosity medium. We numerically quantify and characterize these effects using detailed pore-scale simulations. Our results indicate that in low-porosity complex geological samples, the transition to inertial flows occurs via a distinct mechanism visible as a change in tortuosity and the early formation of vortical structures that evolve with inertial forces and change the structure, dispersion, and shape of preferential flow channels.

\section{Methods}

\begin{figure}[!h]
    \centering
    \includegraphics[width=0.35\linewidth]{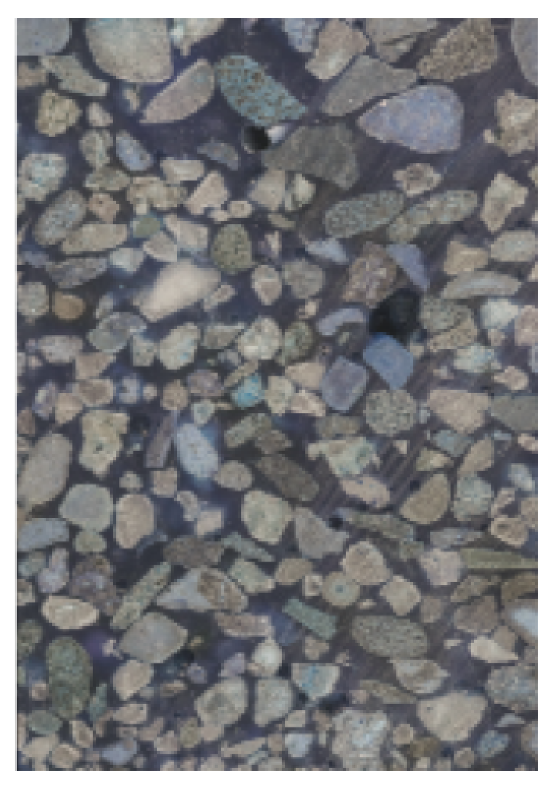}
    \caption{The slice of the intact clastic rock sample filled with synthetic resin (the photo of cross-sectional plane).}
    \label{fig:RGI}
\end{figure}

The reliability and accuracy of numerical simulations of the flow in porous media largely depend on the accuracy of the representation of the geometry of the actual granular medium. The size of the grains, their roughness, and the degree to which they fill the spaces in the rock directly affect the permeability of the rock and tortuosity of the pore channels. A sample of intact clastic rock was taken during the field drilling operations. The pores of the sample were then filled with synthetic resin using a very slow process. Once the resin had hardened, thin slices of clastic rock were prepared from the cylindrical sample, representing different rock cross-sectional planes, and regular, high-resolution photos were taken (see Fig. ~\ref{fig:RGI}). Then, the two‐dimensional pore map was extracted from a 2794×3999 resolution image with porosity \(\phi=0.249\).


For segmentation, we used image processing and created a binary map of pore space and obstacles using a scikit-image library \cite{scikit-image}, extracted clusters of grains using scikit-learn \cite{scikit-learn}, and Scipy \cite{2020SciPy-NMeth}. The clusters were then triangulated with Pyvista \cite{sullivan2019pyvista} to produce geometry in STL format, which served as input for creating a computational domain that was discretized using the cfMesh utility \cite{cfmesh2015}. Our approach generates a structured hexahedral mesh with local refinements in specific regions near the boundaries to increase accuracy.
The steady-state incompressible Navier-Stokes equations governed the fluid flow through the sample, and we solved them at the pore-scale level:
\begin{equation}
\nabla \cdot \mathbf{u}=0, \quad
\rho(\mathbf{u} \cdot \nabla \mathbf{u})=-\nabla p+\mu \nabla^2 \mathbf{u}+\mathbf{f},
\label{eq:nse}
\end{equation}
where $\mathbf{u}$ is the velocity field, $p$ is the pressure, $\rho$ is the fluid density, $\mu$ is the dynamic viscosity, and $\textbf{f}$ is the body force. We used $\rho=1$ and $\mu=2 \cdot 10^{-6}$. We adopt the index notation for the components of vectors, namely, $\mathbf{u}\equiv[u_1,u_2]$, where index $1$ corresponds to the horizontal component and index $2$ corresponds to the vertical component. The computational domain was periodic in the streamwise (vertical) direction. It was created by mirroring the original sample to ensure continuity of the velocity and pressure fields across the domain and omitting the boundary effects known from channel experiments \cite{Koponen97,ko2023prediction}. We solved the Eqs. \ref{eq:nse} in the entire domain, while the post-processing was done for various parameters in the original sample only.
%
%
The results were obtained using the standard finite-volume method implemented in OpenFOAMv.~2212 \cite{jasak2007openfoam}. We used the steady-state simpleFoam solver, which utilizes pressure-velocity coupling and the Semi-Implicit Method for Pressure-Linked Equations (SIMPLE) algorithm \cite{patankar1980numerical}. Numerical steady-state solutions were obtained by ensuring all the residuals below $10^{-6}$.
Simulations were conducted in parallel on a 6-core processor using OpenMPI. To ensure numerical stability and convergence, we set the pressure under-relaxation to 0.6 and the velocity under-relaxation to 0.9, following established practices that offer a good trade-off between stability and solution accuracy \cite{jasak2007openfoam}.

\section{Results}

\begin{figure}[!h]
    \centering
    \includegraphics[width=1\linewidth]{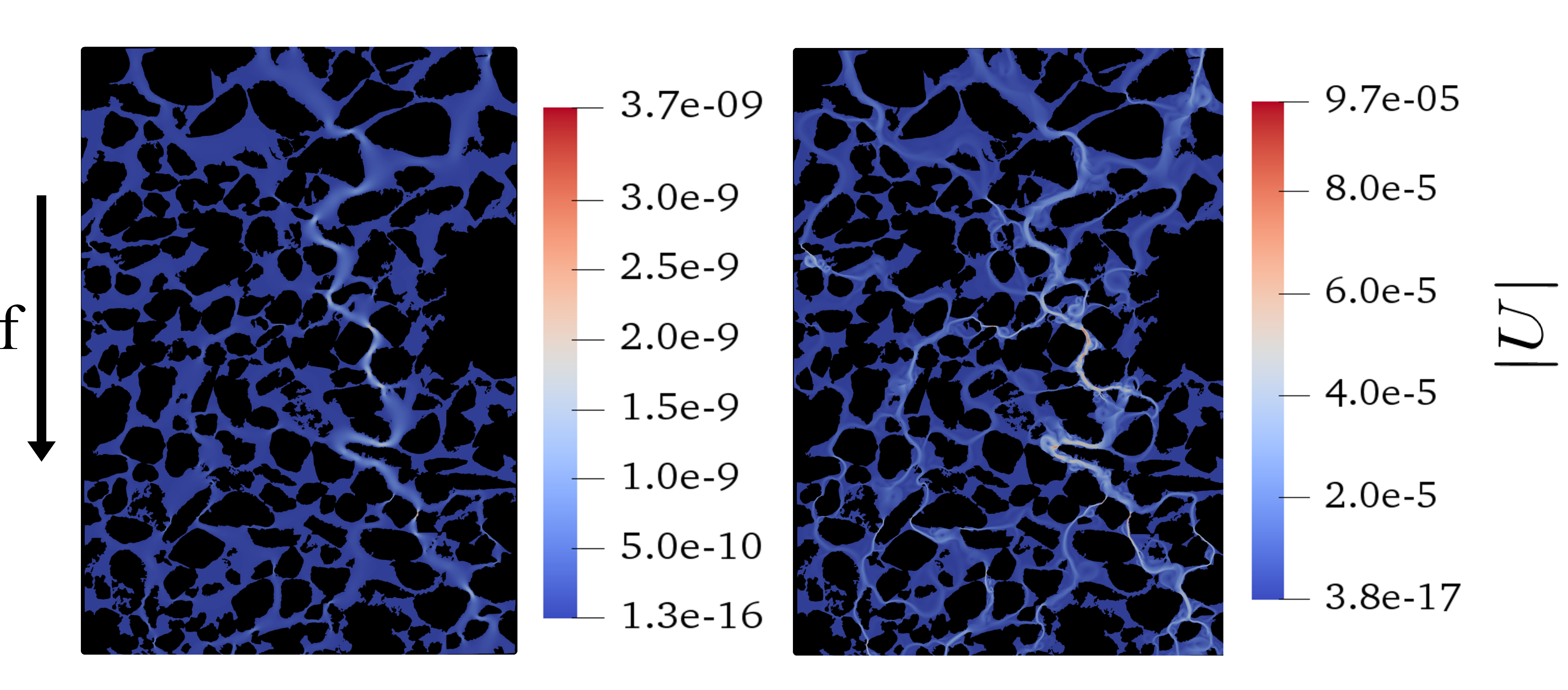}
    \caption{The magnitude of the pore-space velocity field at $Re=0.03$ (left) and $Re=9.62$ (right). The arrow indicates the direction of the body force driving the flow.}
    \label{fig:velocity-field}
\end{figure}

The pore space is filled with fluid and subjected to an external body force to drive the flow. The force $\textbf{f}$ starts from extremely low value $\mathbf{f}= (0,-2 \times 10^{-15}) [kg/m^2s^2]$ and increases, which results in changes in the Reynolds number, which is defined as $Re=|\mathbf{U}| \cdot L/\nu$, where $L$ is the width of the sample and $\mathbf{U}$ is the superficial velocity (Eq. ~\ref{eq:darcy}) and $\nu$ is the kinematic viscosity. The maximum Reynolds number was set to $10^3$. Although local velocities in the narrow channels may yield a higher local $Re$, all SIMPLE runs remained numerically stable. 
The resulting velocity fields at two selected Reynolds numbers are shown in Fig.~\ref{fig:velocity-field}. These results demonstrate the formation of a preferential channel at low Reynolds numbers (Fig.~\ref{fig:velocity-field}, left), primarily governed by viscous forces and increased momentum dispersion in high-Reynolds-flow (Fig.~\ref{fig:velocity-field}, right), resulting in dispersed preferential channels where the maximum velocities are almost four orders of magnitude higher. A closer look at the flow field reveals that the vortex formation process is visible in the velocity field structure. 

\begin{figure}[!h]
    \centering
    \includegraphics[width=1\linewidth]{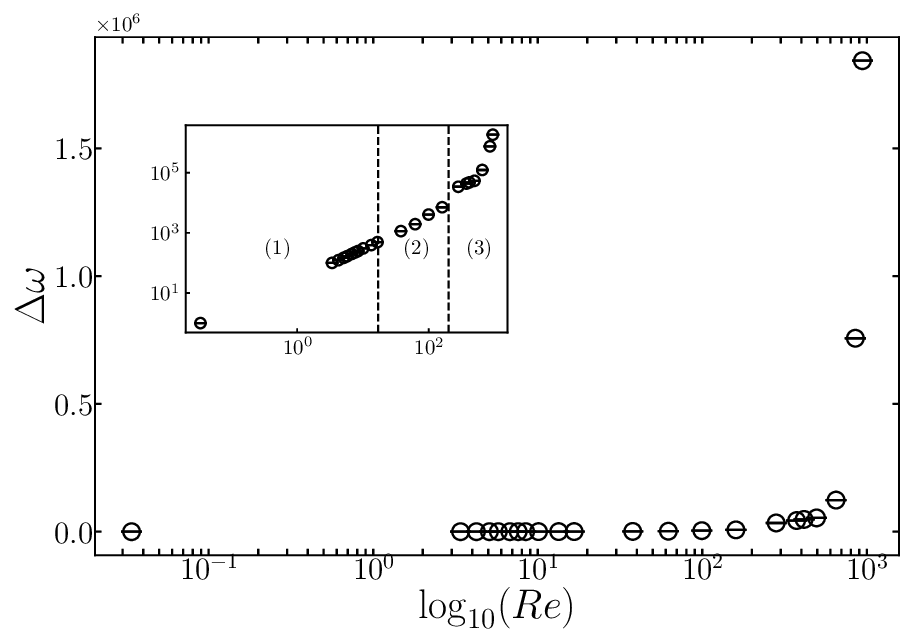}
    \caption{Mean values of $\Delta\omega$—the sum of the absolute difference in vorticity magnitude between the upper and lower halves of the domain—plotted versus $Re$, with vertical bars showing the full min–max range at each $Re$. The inset presents the same data on a log–log scale and highlights three regimes: (1) the power law ($\Delta\omega \propto Re^p$), (2) slightly faster-than-exponential growth, and (3) a rapid “explosive” increase. }
    \label{fig:assym}
\end{figure}

Next, we quantified the asymmetry in the velocity field by investigating the vorticity ($\omega=\frac{\partial u_2}{\partial x}- \frac{\partial u_1}{\partial y}$) field over the entire computational domain (see Fig.~\ref{fig:assym}), which is divided into an upper (main) domain and its corresponding mirrored (buffer) domain. We compared the vorticity magnitudes at the corresponding nodes between these two regions. We defined the origin of the coordinate system at the bottom of the original sample.
For each node $\mathbf{x_j}\equiv(x_j,y_j)$ in the upper (original) domain, we identified its mirror node $\mathbf{x_j}'\equiv(x_j,2H-y_j)$ in the lower (mirror) domain, where $H$ is the size of one-half of the domain along the streamwise direction. Then, we calculated the sum of the absolute difference in vorticity magnitudes as
\begin{equation}
\Delta\omega
= \sum_{j=1}^{N}
  \bigl|\,
    \lvert\omega(\mathbf{x}_j)\rvert
    - \lvert\omega(\mathbf{x}_j')\rvert
  \bigr|
  \quad\text{with }\{\mathbf{x}_j\}_{j=1}^N=\Omega_U,
\end{equation}
where $\Omega_U$ denotes the set of nodes in the upper domain. As shown in Fig.~\ref{fig:assym}, $\Delta \omega$ remained low at low Reynolds numbers, indicating that the flow remained symmetric in the Darcy regime. However, as the Reynolds number increases, the asymmetry of the flow field increases rapidly due to inertial effects. The main plot illustrates this transition, while the log-log inset confirms that the asymmetry grows monotonically across the entire range of Reynolds numbers, with a pronounced increase at relatively high Reynolds numbers. This increase in asymmetry highlights the emergence of complex flow structures. This suggests that factors other than pore geometry (symmetric in our case) play a crucial role in channel formation at high Re values.

\begin{figure}[h!]
    \centering
    \includegraphics[width=1\linewidth]{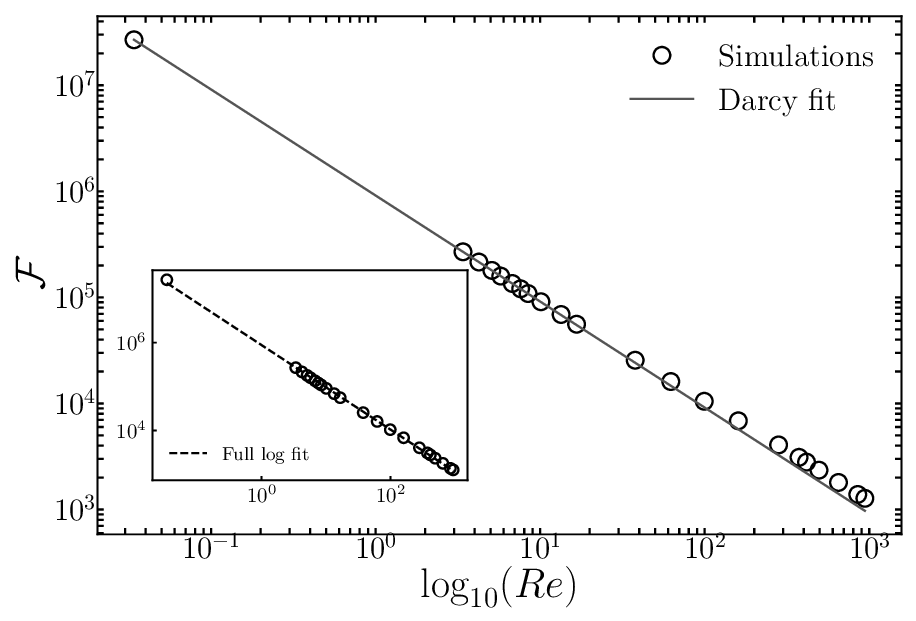}
    \caption{Generalized friction factor $\mathcal{F}$ versus \( \log_{10}(Re) \), showing simulation data and a linear fit to the first four data points (solid line). The dashed line (inset) illustrates the full-range log-log fit.}
    \label{fig:fric}
\end{figure}


To assess the influence of inertia on the structure and its correlation with the asymmetry of the flow, we computed the friction factor \cite{andrade1999inertial} as follows:
\begin{equation}
    \mathcal{F}= \frac{\textbf{f} L}{\rho \langle |\mathbf{u}|\rangle^2}.
\end{equation}
In a highly porous medium, the friction factor shows an apparent deviation at high velocities \cite{andrade1999inertial, sniezek2024inertia} and indicates a transition from Darcy to nonlinear Forchheimer flows, where inertial forces become significant. However, in our simulations, this transition was not immediately evident (Fig.~\ref{fig:fric}). This is due to the low porosity and the dominant effect of the pore geometry guiding the flow through the system. We noticed that, if the numerical fit of Darcy, Eq.~\ref{eq:darcy} is taken over all data points, the friction factor follows the fit for the entire range of Reynolds numbers (see inset in Fig.~\ref{fig:fric}). Therefore, the flow at the highest Reynolds numbers may be mistakenly considered to lie in the Darcy regime.

\begin{figure}[!h]
    \centering
    \includegraphics[width=1\linewidth]{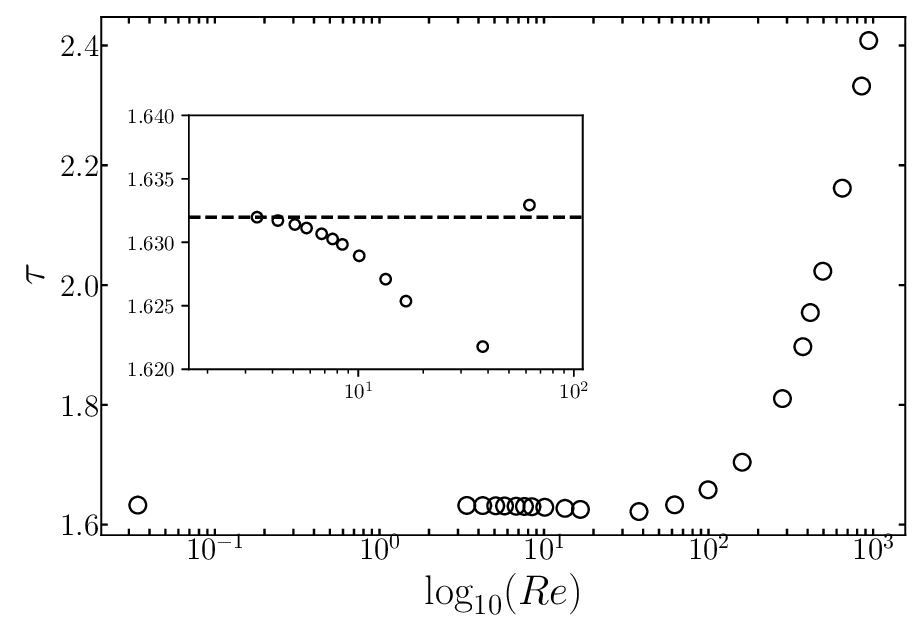}
    \caption{The tortuosity $\tau$ calculated using Eq. ~\ref{eq:tau} with increasing $Re$. The inset shows the first 14 data points, excluding the minimum $Re$.}
    \label{fig:tau}
\end{figure}

To quantify the inertial effects visible in the velocity fields, as shown in Fig.~\ref{fig:velocity-field}, we analyzed flow tortuosity $\tau$, a dimensionless number representing the relative elongation of the fluid path compared to the sample length. We computed it from the pore-scale velocity field as:
\begin{equation} \label{eq:tau}
\tau = \frac{\langle |\mathbf{u}| \rangle}{\langle u_2 \rangle} 
\end{equation}
where $\langle u_2 \rangle$ is the average streamwise velocity component and $\langle |\mathbf{u}| \rangle$ is the average magnitude of the velocity \cite{Duda11}. 
Our results (Figs.~\ref{fig:tau}) indicate that the tortuosity remains constant at low $Re$ and exhibits a relatively small decrease (see inset of Fig.~\ref{fig:tau}), reaching a minimum around $Re=30$. This observation is consistent with our previously reported results for 3D, highly porous media \cite{sniezek2024inertia}, where such a decrease in tortuosity is also noticeable. However, above this minimum, $\tau$ increases with Re, which has not been observed before. 
The non-monotonous rise of $\tau$ with $Re$ suggests that the two competing mechanisms influence the tortuosity as the Reynolds number increases, and their relative importance depends on factors such as the flow speed, porosity, and geometric characteristics of the porous structure.
%

\begin{figure}[!h]
    \centering
    \includegraphics[width=1\linewidth]{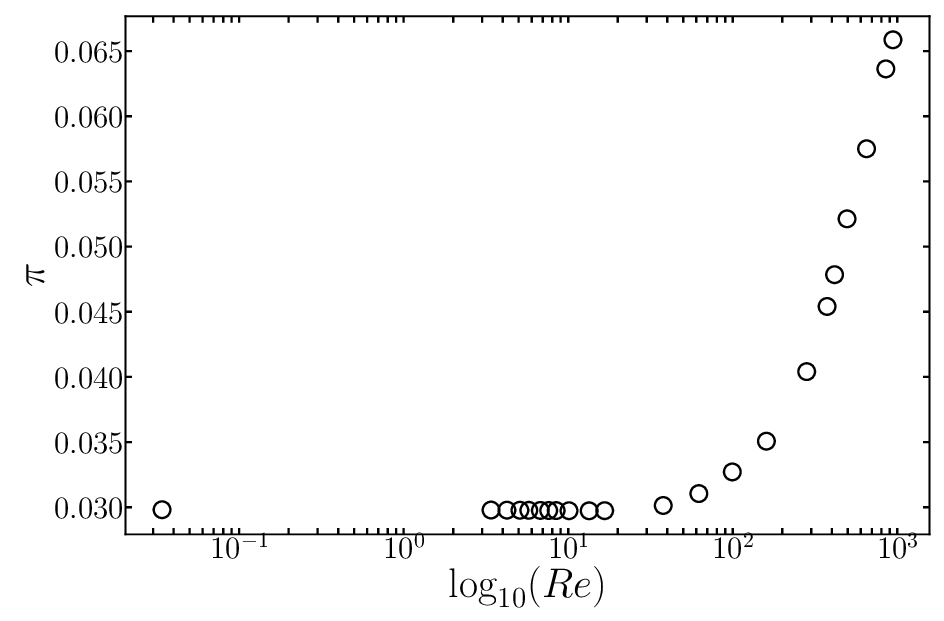}
    \caption{The growth of participation number $\pi$ with $Re$.}
    \label{fig:pi}
\end{figure}

To understand the mechanism of the tortuosity changes, we investigated the fluid's kinetic energy distribution in the pore space. We calculated the participation number $\pi$, which quantifies the localization of kinetic energy distribution \cite{andrade1999inertial}. It is defined as:
\begin{equation}
\pi\equiv \left(V \sum_{i=1}^n q_i^2 V_i\right)^{-1},     
\end{equation}
where $V_i$ is $i$-th cell, and $q_i=\frac{e_i}{\sum_{j=1}^n e_j V_j}$ with $e_i \propto \boldsymbol{u}_i^2$ \cite{andrade1999inertial}. The relationship between participation number and the Reynolds number is shown in Fig.~\ref{fig:pi}. 
At low Reynolds numbers, $\pi$ remains low and nearly constant, indicating the concentration of kinetic energy in a small fraction of pores and that we are in the Darcy regime. This is consistent with the visual inspection of the velocity field shown in Fig.~\ref{fig:velocity-field}, where at a low Reynolds number, the flow is primarily through a single channel that emerges in the system. We find that $\pi$ increases monotonically with $Re$, indicating that kinetic energy becomes more uniformly distributed across the pore space as inertial effects increase. This increase indicates that at high $Re$, the velocities outside the main channels and existing vortical cores increase even faster than those within them, revealing the formation of new flow pathways and a broader redistribution of kinetic energy. This behavior sheds more light on the tortuosity behavior shown in Fig.~\ref{fig:tau}, where the rapid rise may now be associated with an increasing number of vortices in the flow and the definition of tortuosity that was used. It is also notable that our results in the highly nonlinear regime, show $\pi\propto \tau$, which may be interesting to investigate in the future.
 
\begin{figure}[]
    \centering
    \includegraphics[width=1\linewidth]{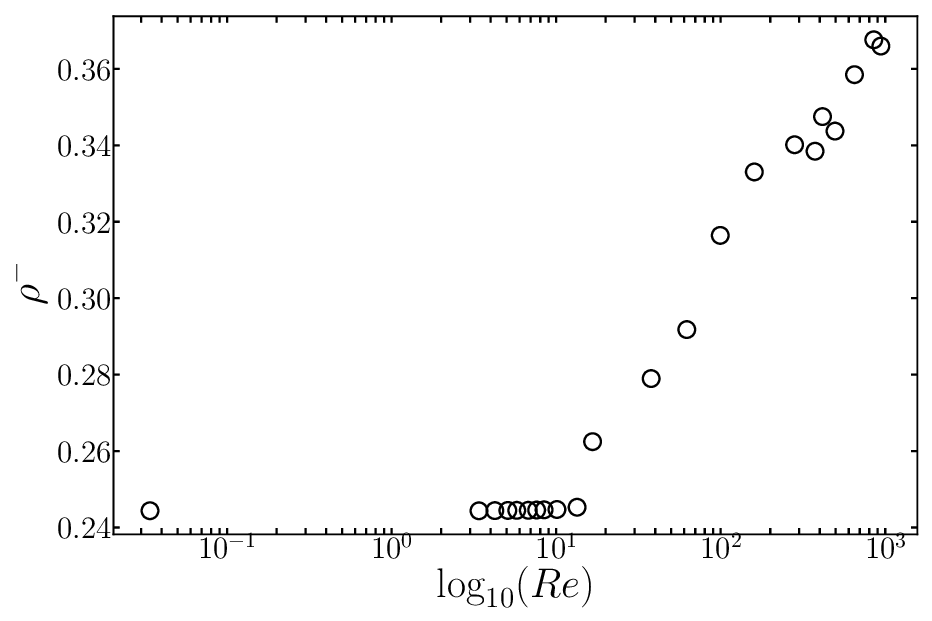}
    \caption{$\rho^{-}$ versus $Re$.}
    \label{fig:rho}
\end{figure}

\begin{figure}[]
    \centering
    \includegraphics[width=0.75\linewidth]{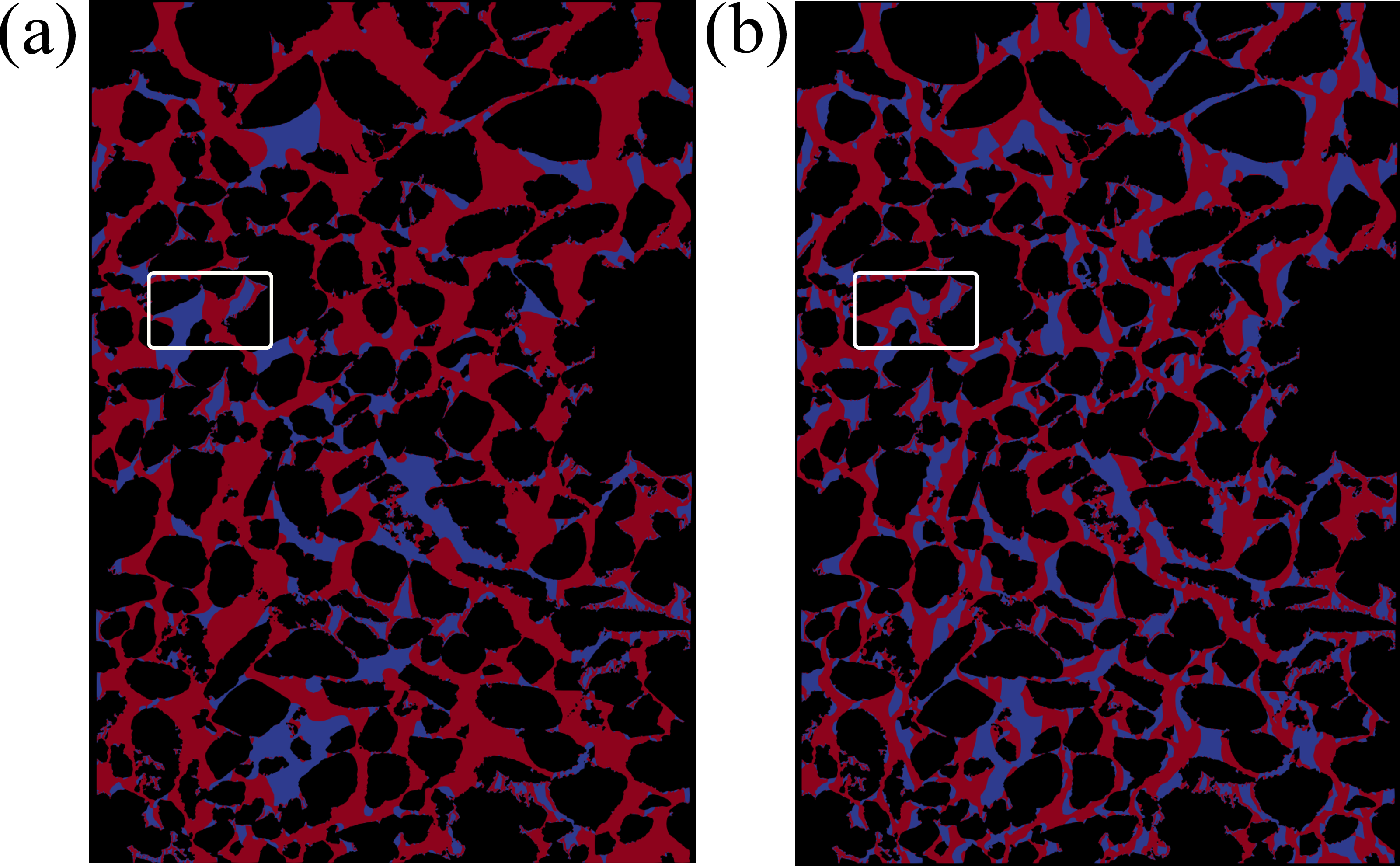}  
    \caption{The maps of $f(u_2)$ (Eq. ~\eqref{eq:velocity_indicator}) for (a) $Re=0.03$, and (b) $Re=940.94$. Red (dark in grayscale) indicates $f(u_2)=1$, whereas blue (light in grayscale) indicates $f(u_2)=0$. The white frame in each plot indicates the areas shown at the magnification in Fig.~\ref{fig:velocity_indicator}.}
    \label{fig:velocity_indicator}
\end{figure}

\begin{figure}[]
    \centering
    \includegraphics[width=0.75\linewidth]{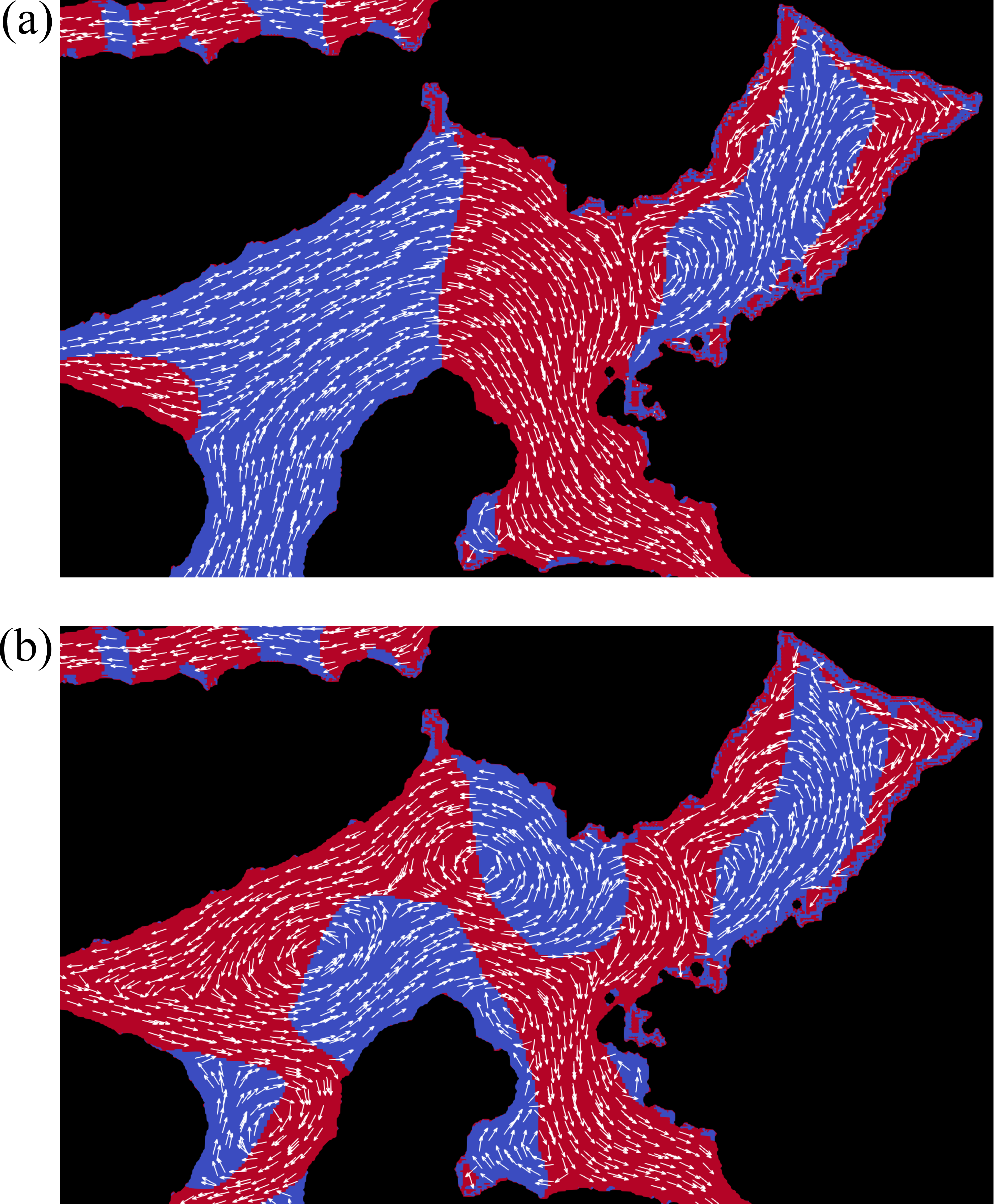}  
    \caption{The magnification of particular regions of the maps of $f(u_2)$ (Eq.~\eqref{eq:velocity_indicator}) as shown in Fig.~\ref{fig:velocity_indicator}. Vectors represent the local direction of the velocity field. For clarity, vectors close to the solid boundaries are omitted.}
    \label{fig:velocity_indicator_zoom}
\end{figure}

The steep rise in $\pi$ is related to the reduction in preferential channeling accompanied by the emergence of vortices and flow recirculation zones in the pore space. To show this, we computed the parameter $\rho^{-}$ (see Fig. \ref{fig:rho}), introduced by us in \cite{sniezek2024inertia}, which quantifies the volume fraction of the pore-space containing negative streamwise velocity. It is defined as: 
\begin{equation}
\rho^{-}= \frac{1}{V} \sum_{i=1}^n f\left(u_{2, i}\right) V_i,
\label{eq:rho-minus}
\end{equation}
where $u_{2,i}(\mathbf{r})$ is the streamwise velocity component, and $V_i$ represents the volume of $i$-th cell. The function $f\left(u_2\right)$ is an indicator function defined as
\begin{equation}\label{eq:velocity_indicator}
f\left(u_2\right)=\left\{\begin{array}{cc}
1, & u_2<0 \\
0, & \text { otherwise }.
\end{array}\right.
\end{equation}
As shown in Fig.~\ref{fig:rho}, at low Reynolds numbers, $\rho^{-}$ is not zero and is constant ($\rho^{-}\approx0.24$), indicating the presence of flow reversal in the Darcy regime. We investigated this phenomenon (see Fig.~\ref{fig:velocity_indicator} and Fig.~\ref{fig:velocity_indicator_zoom}). We noticed that this occurs because of the combined effect of a small fraction of vortices appearing at the lowest Re and the dominant geometry-driven backward flow. In contrast, $\rho^{-}$ investigated in a highly porous system \cite{sniezek2024inertia} was nearly $0$ at a low Re. Moreover, as $Re$ increases, $\rho^{-}$ rises rapidly around $Re\approx 10$, earlier than the rise in $\tau$ is observed. 
%
%

Our analysis of $\rho^{-}$ (Fig.~\ref{fig:rho}) suggest that vortices appear earlier than significant changes in tortuosity and before a substantial increase in the energy distribution. To understand this and describe the physical mechanism behind the $\rho^{-}$ increase, we visualized the spatial map distribution of the backward-flow indicator $f(u_2)$. This highlights the regions of reverse flow within the pore space (Fig.~\ref{fig:velocity_indicator}, and zoomed regions in Fig.~\ref{fig:velocity_indicator_zoom} for two representative Reynolds numbers). Surprisingly, at a low Reynolds number ($Re=0.03$), isolated flow pockets of non-zero $\rho^{-}$ appear and confirm the presence of backward flow and localized vortices. As the Reynolds number increased, the reverse-flow regions shrank in favor of the dominant vortex component (Fig.~\ref{fig:velocity_indicator}, right panel), forming numerous neighboring recirculation zones. For a relatively high $Re=940.94$, a complex and widely dispersed vortex structure system was visible throughout the entire pore space. The extensive, more dispersed backward flow regions, now in high Re, visible primarily as vortices, correlate with the sharp rise in  $\rho^{-}$ observed in Fig.~\ref{fig:rho}, as well as the increasing trends in $\tau$ and $\phi$.

\begin{figure}[h!]
    \centering
    \includegraphics[width=1\linewidth]{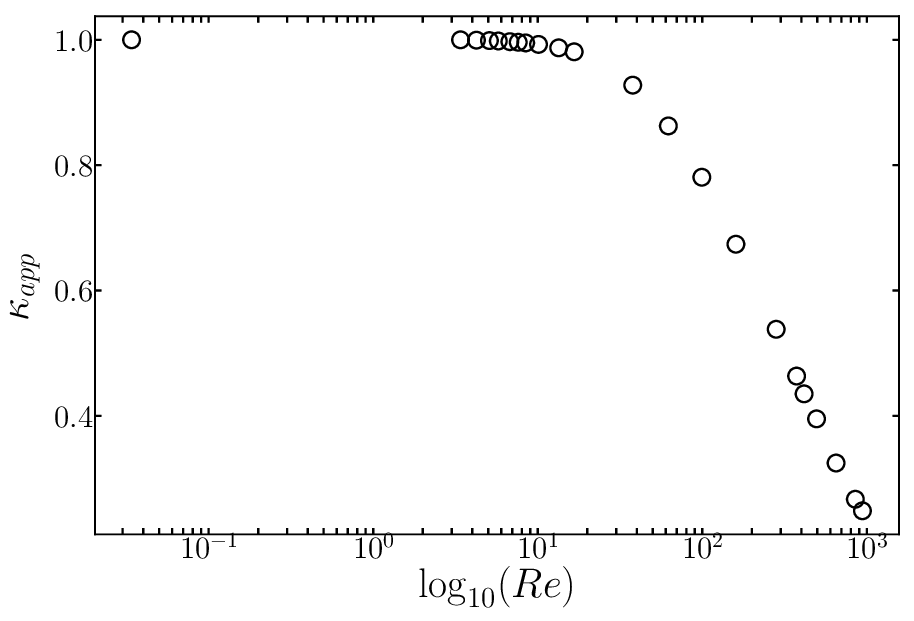}
    \caption{The normalized apparent permeability $\kappa_{app}$ versus $Re$. The inset plot is $\kappa$ vs. $\tau$.}
    \label{fig:kappa}
\end{figure}

\begin{figure}[h!]
    \centering
    \includegraphics[width=1\linewidth]{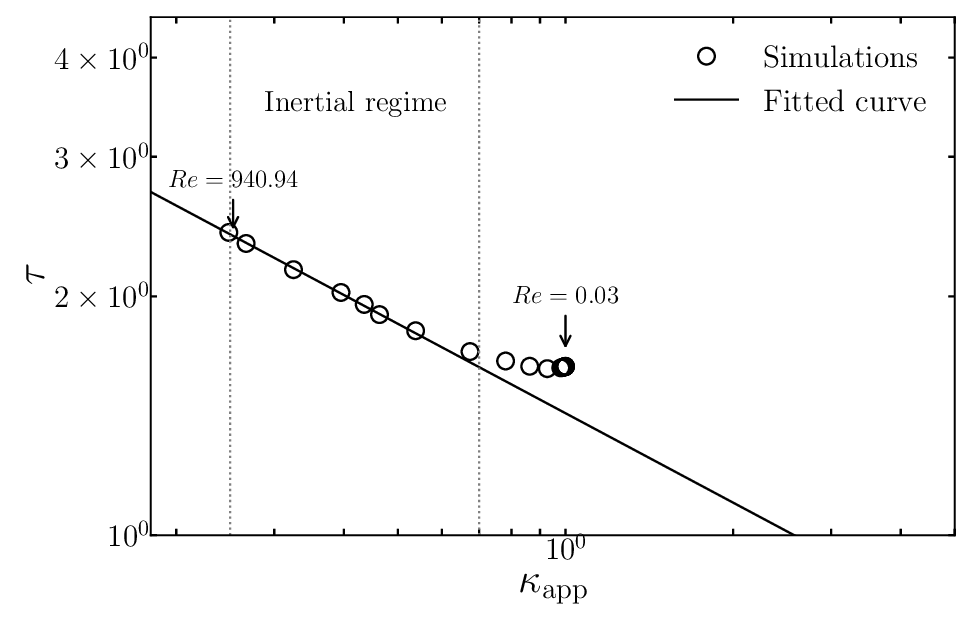}
    \caption{Correlation between tortuosity $\tau$ and the apparent permeability $\kappa_{app}$ measured in the flow. The numerical fit of the power law relation was taken in the inertial regime and drawn as a solid line. The dashed lines guide the eye to distinguish the inertial regime of the flow roughly.}
    \label{fig:tau_kappa}
\end{figure}

Finally, we investigate the apparent permeability $\kappa_{app}$ by computing its value for all $Re$ directly from the Darcy relation (even though the permeability $\kappa$ is defined initially for the Darcy regime, we use the apparent permeability definition \cite{BarreeConway2004,BalhoffWheeler2009}). In Fig.~\ref{fig:kappa}, we present the variation in $\kappa_{app}$ (normalized by its value at the lowest $Re$) with $Re$. At low Reynolds numbers, $\kappa_{app}=\kappa$ and remains nearly constant, indicating that the flow is likely dominated by viscous effects, resembling the dominant role of geometry on permeability values. However, $\kappa_{app}$ decreased in the nonlinear regime. 
It is evident from our results, that permeability and tortuosity are correlated with an inertia-driven increase in tortuosity, leading to a decrease in permeability.
%
To quantify this, we fitted the power law $\kappa_{app}\propto \tau^{-b}$ and found $b\approx 1/3$ (see Fig.~\ref{fig:tau_kappa}). This behavior aligns with the results of the impact of tortuosity on permeability in the flow observed in the Darcy regime for systems of varying porosities reported previously \cite{Koponen97}. Intriguingly, in the inertial regime, the appearance of vortices accelerates the flow in narrower, more tortuous channels, effectively accelerating the macroscopic flow by reducing shear stresses along the pore walls. However, this process still leads to a net decrease in apparent permeability.

\section*{Conclusions}

We have shown that low-porosity systems exhibit a different transition mechanism from a low Reynolds number (Darcy flow) to a high Reynolds number (non-Darcy flow) in a low-porous medium than in a highly porous medium \cite{sniezek2024inertia}. We found that the complex structures of the pore geometry led to the emergence of vortices even at low Reynolds numbers. Moreover, the study of $\rho^{-}$ indicates the backward flow induced by geometry. To understand this, we investigated tortuosity and observed its non-monotonic growth. We associate this with two physical mechanisms: straightening the flow streamlines due to increased momentum localized in the channels with a more dispersed flow structure in general (lowering $\tau$), and the appearance of vortices (increasing $\tau$). Finally, we compared the tortuosity with the apparent permeability and determined the power law that holds in the inertial regime.

\section*{Acknowledgment}

We are grateful to Jos\`e S. Andrade Jr. for insightful
discussion and comments during the early stages of the
work. Funded by National Science Centre, Poland under the OPUS call in the Weave programme 2021/43/I/ST3/00228.
This research was funded in whole or in part by National Science Centre (2021/43/I/ST3/00228). For the purpose of Open Access,
the author has applied a CC-BY public copyright licence to any Author Accepted Manuscript (AAM) version arising from this submission.
M. Matyka acknowledges the financial support from the Slovenian Research And Innovation Agency (ARIS) research core funding No. P2-0095.

\bibliographystyle{elsarticle-num}

\end{document}